# A NEW GRAPHICAL MICROBURST GUIDANCE PRODUCT

Kenneth L. Pryor
Center for Satellite Applications and Research (NOAA/NESDIS)
Camp Springs, MD

## 1. Introduction

A new microburst graphical guidance product has been developed that employs data from the Rapid Update Cycle (RUC) model. Prototypical conditions for microbursts include a steep temperature lapse rate and decreasing humidity with decreasing height in the boundary layer. Thus, the graphical guidance product incorporates boundary layer temperature lapse rate, vertical relative humidity difference, and precipitable water. These parameters have been noted as important factors in initiating and sustaining a convective downdraft. The new guidance product has demonstrated effectiveness in indicating favorable conditions for downbursts over the northern Chesapeake Bay region of Maryland. This paper will introduce the RUC graphical microburst product by presenting case studies that highlight effective performance.

## 2. Methodology

The developmental microburst guidance product is generated by Man computer Interactive Data Access System (McIDAS-V) software in which fields of interest for microburst potential are overlain to produce a composite image. The product image is derived from RUC model datasets in gridded binary (GRIB) format from NOAA National Operational Model Archive & Distribution System (NOMADS). The interest fields include temperature lapse rate, vertical humidity difference, and precipitable water. In most cases, lapse rate and vertical humidity difference are calculated for the layer between 1000 and 850mb, assuming a cloud base height near the 850-mb level. Caracena and Flueck (1988) and Srivastava (1985) noted that the sub-cloud temperature lapse rate and dewpoint depression difference (proxy for relative humidity difference) are important predictors of microburst activity. Temperature lapse rate is represented as a color-shaded graphic while humidity difference and precipitable water are represented by color contouring. In McIDAS-V, radar imagery (i.e. reflectivity and velocity) can be overlain on the microburst graphic product, thereby providing a source of "truth" in the validation process. For each downburst event, the microburst product image and associated values for lapse rate, vertical humidity difference, and precipitable water are compared to proximate surface observations of convective storm-generated high wind gusts by National Ocean Service (NOS) Physical Oceanographic Real-Time System (PORTS) stations. The temporal resolution of NOS data, six minutes, is well-suited for downburst observation. Next Generation Radar (NEXRAD) and Terminal Doppler Weather Radar (TDWR) base reflectivity imagery from National Climatic Data Center (NCDC) were utilized to verify that observed wind gusts were associated with downbursts and not associated with other types of convective wind phenomena (i.e. gust fronts). Another application of the radar imagery was to infer microscale physical properties of downburst-producing convective storms. Radar velocity imagery was utilized to estimate the magnitude of downbursts where surface observation data was insufficient.

## 3. Case Studies

### 3.1 Case 1: March 2004 Chesapeake Bay Downbursts

Analysis of RUC model data has provided new results from the study of the historic March 2004 Chesapeake Bay downburst event. This event was associated with the Baltimore water taxi accident that occurred during the afternoon of 6 March 2004. Computation and analysis of RUC-derived downdraft instability parameters, including temperature lapse rate, vertical relative humidity difference, and precipitable water, revealed local maxima in proximity to downburst occurrence about one hour prior.

The first downburst resulted in the capsize of the "Lady D" in the Baltimore Harbor. As displayed in Figures 1 and 2, and noted in Table 1, the stronger downburst recorded at Tolchester Beach PORTS station was associated with higher values of all the listed parameters. In general, the stronger downburst was associated with a steeper sub-cloud temperature lapse rate and a larger vertical humidity difference below 850mb, and higher storm precipitable water content. In accordance with findings of Srivastava (1985), downbursts were associated with sub-cloud lapse rates greater than 8.5 K/km. This suggests that sub-cloud evaporational cooling in a more well-mixed boundary layer and precipitation loading were factors in the generation of downdraft instability and resulting strong downbursts. These conditions, more typically found over the Great Plains during the warm season, were effectively indicated by RUC analysis-derived parameters about one hour prior to the first downburst occurrence near the Baltimore Harbor.

### 3.2 Case 2: July 2009 Downbursts

During the late afternoon and evening of 25 July 2009, a quasi-linear cluster of convective storms developed over the Maryland Piedmont and tracked eastward toward the Chesapeake Bay. Decaying convective storm cells triggered the development of new cells along the leading edge of the cluster over the Baltimore Harbor and Western Shore of the Chesapeake Bay. Strong convective wind gusts were recorded by PORTS stations at the Francis Scott Key (FSK) Bridge shortly after 0000 UTC 26 July. 0000 UTC RUC graphical guidance indicated favorable conditions for downbursts over the Baltimore Harbor with temperature lapse rates greater than 8 K/km and precipitable water values near 48mm (2 inches). Figure 3 compares graphical guidance with a RUC sounding profile over the Baltimore Harbor near the time downburst occurrence at 0012 UTC.

A favorable downburst environment is illustrated in Figure 3. The juxtaposition of an unstable temperature lapse rate and large precipitable water values near 2 inches over the Baltimore Harbor suggests that precipitation loading was a major forcing mechanism for strong convective winds. The RUC sounding profile displayed a modified "inverted V" profile with a well-developed mixed layer below 800 mb that promoted sub-cloud evaporational cooling and the generation of negative buoyancy. Radar imagery from Baltimore-Washington International Airport (BWI) TDWR confirmed this condition with maximum storm reflectivity indicated near

45 dBZ.  The downburst-producing convective storm tracked over the Baltimore Harbor between 0009 and 0015 UTC 26 July.  Downburst occurrence was confirmed by PORTS observations at the FSK Bridge in Figure 4 and radar imagery in Figure 3 as the parent storm moved over the Baltimore Harbor.  Based on the relationship established in Srivastava (1985), lapse rates greater than 8 K/km in combination with radar reflectivity near or above 45 dBZ yields a high likelihood for microbursts.

As the storm cell moved over the harbor near the FSK Bridge, PORTS observations in Figure 4 were consistent with downburst occurrence.  A sharp peak in wind gust speed to 36 knots, a 2 mb pressure jump, and sharp temperature decrease indicated downburst occurrence between 0010 and 0015 UTC.  Wind gusts in excess of 35 knots posed a hazard to both marine and ground transportation in the vicinity of the FSK Bridge.  As suggested in the product image and sounding in Figure 3, this downburst event highlighted the importance of precipitation loading and sub-cloud evaporational cooling in a conditionally unstable environment in the generation of strong convective winds.

### 3.3  Case 3:  November 2009 Downburst

The new guidance product demonstrated effectiveness in indicating favorable conditions for downbursts over the northern Chesapeake Bay region during the afternoon of 5 November 2009 in which a convective storm produced a strong wind gust of 38 knots at Tolchester Beach, Maryland. The 1800 UTC RUC graphical microburst product indicated high downburst risk in proximity to Tolchester Beach.

Figure 5 compares the new RUC graphical guidance microburst product to a corresponding RUC sounding over Tolchester Beach, Maryland at 1800 UTC, 5 November 2009. At 1800 UTC, the image product showed a large area with steep boundary layer lapse rates (orange shading), greater than 8.5 K/km, over central and northeastern Maryland. The product image also displayed a local maximum in vertical humidity difference over northern Maryland. The highest microburst risk was indicated where the highest vertical humidity difference was co-located with steep lapse rates in the 850-1000mb layer. Overlying radar reflectivity imagery from BWI TDWR at 2235 UTC displayed a downburst-producing convective storm as a bow echo (Przybylinski 1995) near Tolchester Beach. At 2242 UTC, the Tolchester Beach PORTS station recorded a wind gust of 38 knots as illustrated in Figure 6. The corresponding RUC sounding profile echoed favorable conditions for downbursts in the Tolchester Beach area with the presence of a 5000-foot deep mixed layer and steep temperature lapse rate below 850mb. Although radar reflectivity with this storm was not impressive (25-40 dBZ), the steep lapse rate was a major contributor to downdraft instability.

### 4. Summary and Conclusions

As demonstrated in these case studies, the RUC graphical guidance product, visualized by McIDAS-V software, effectively highlighted regions favored for strong convective winds. In accordance with the findings of Srivastava (1985), downburst activity was associated with sub-cloud lapse rates greater than 8.5 K/km and independent of storm radar reflectivity with high reflectivity associated with the March 2004 and July 2009 events and low reflectivity associated with the November 2009 event. In general, downburst activity was associated with environments where precipitation loading and sub-cloud evaporational cooling fostered strong convective downdrafts.  The ability to overly radar reflectivity, especially from TDWR, on the composite guidance product shows the utility of the guidance in the downburst nowcasting process.

| Time (UTC) | Location | Wind Gust (kt) | Lapse Rate (K/km) | dRH(%) | PW (mm) |
| --- | --- | --- | --- | --- | --- |
| 2050 | Baltimore Harbor | 35 to 45 | 8.6 | 16 | 25 |
| 2118 | Tolchester Beach | 48 | 8.9 | 17 | 27 |

Table 1.  Two strong downbursts that occurred in the upper Chesapeake Bay region between 2050 and 2120 UTC 6 March 2004 and associated RUC-derived microburst parameters from 2000 UTC.

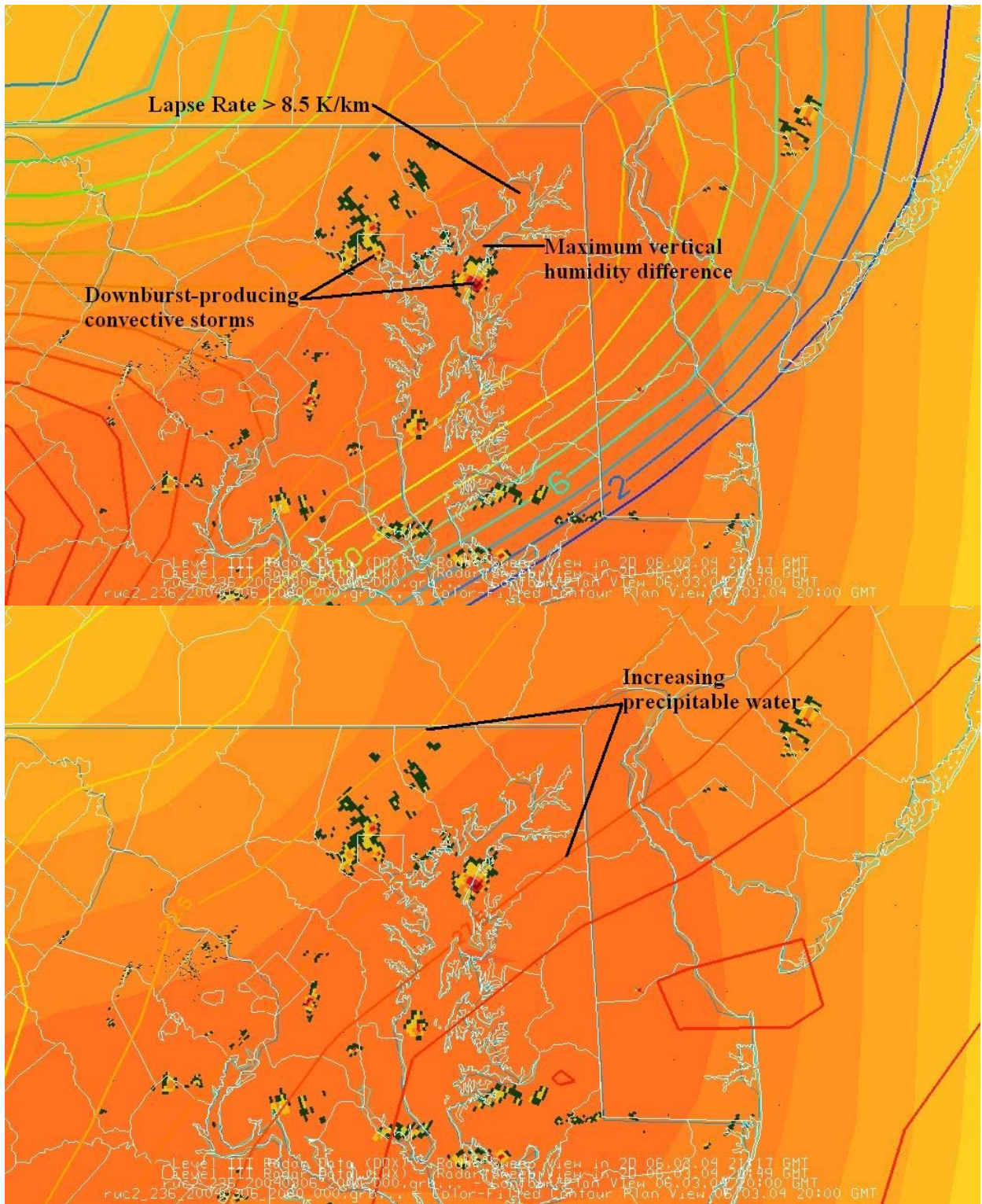

Figure 1. RUC derived temperature lapse rate and vertical humidity difference (dRH, top), and precipitable water (PW, bottom) at 2000 UTC 6 March 2004 with overlying NEXRAD radar reflectivity .

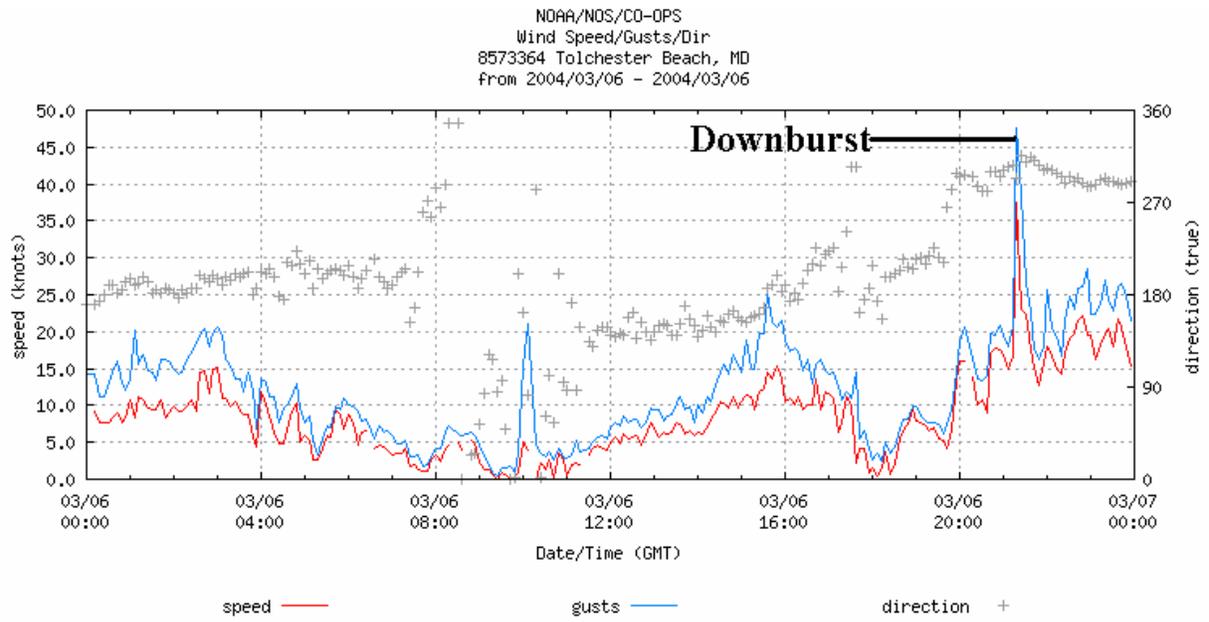

Figure 2. Wind histogram from Tolchester Beach PORTS station.

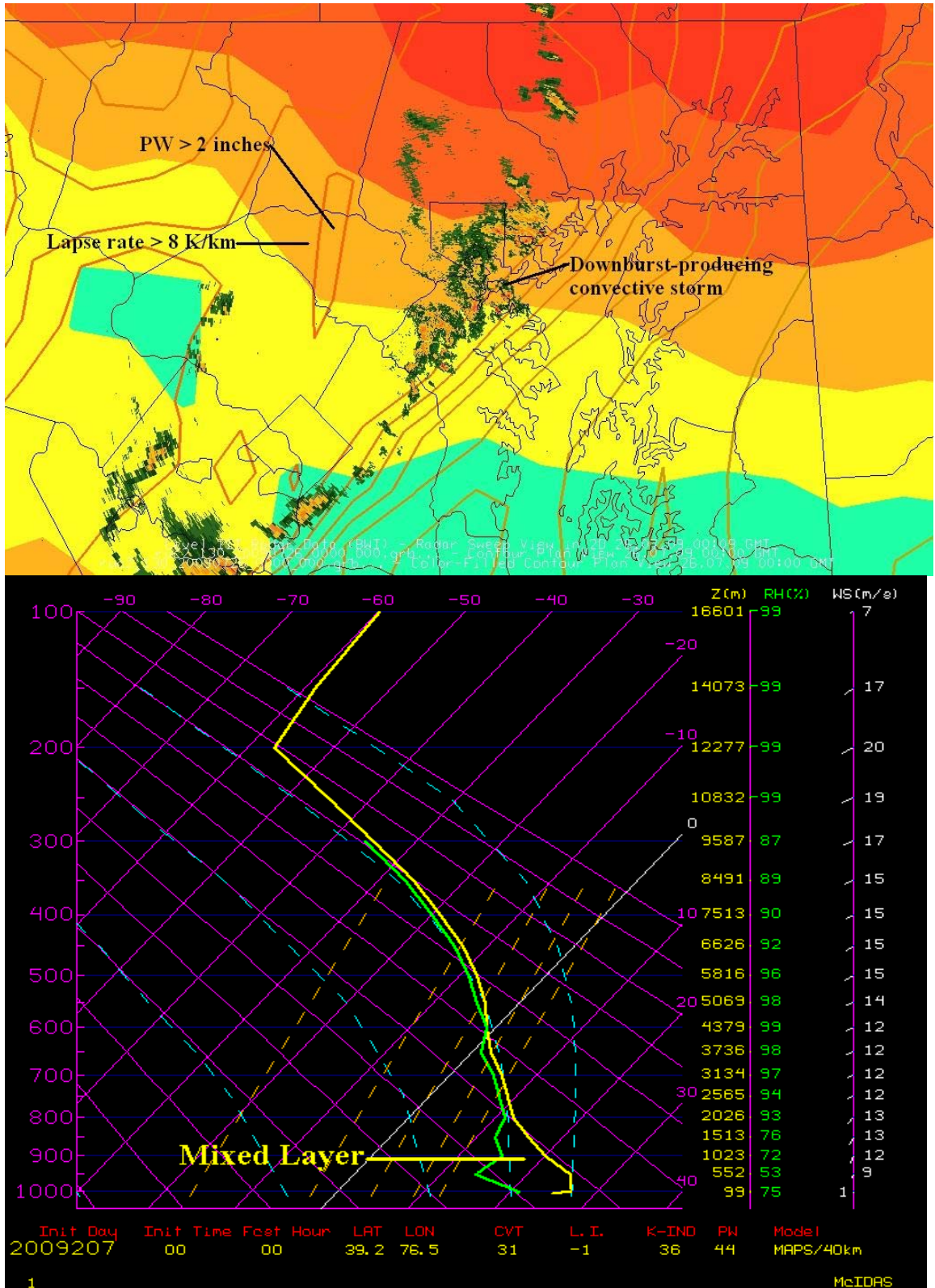

Figure 3. RUC model-derived graphical microburst product at 0000 UTC 26 July 2009 with overlying radar reflectivity from BWI TDWR (top) and corresponding RUC sounding profile over the Baltimore Harbor.

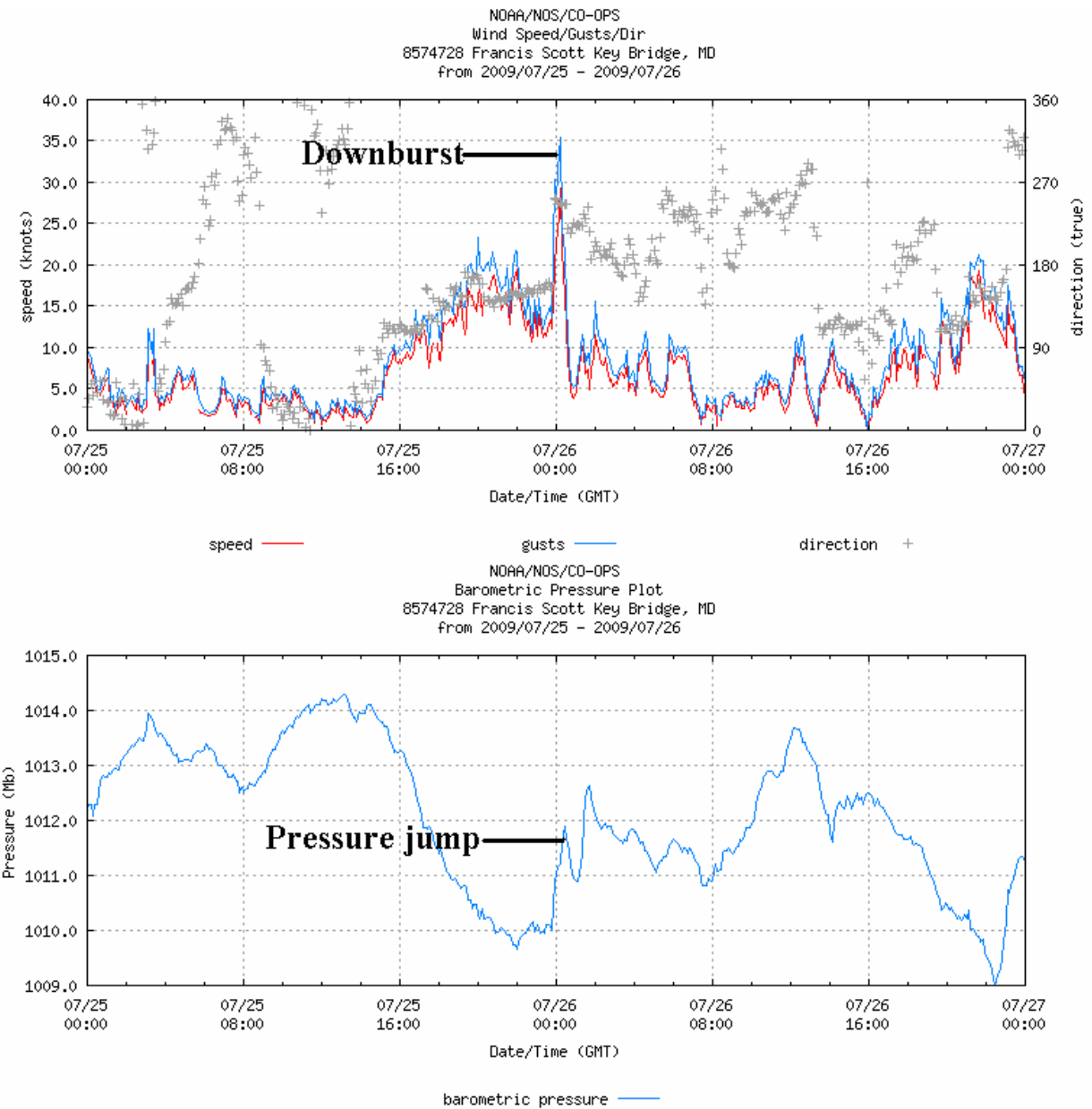

Figure 4. Histograms of wind (top) and barometric pressure (bottom) from FSK Bridge PORTS station.

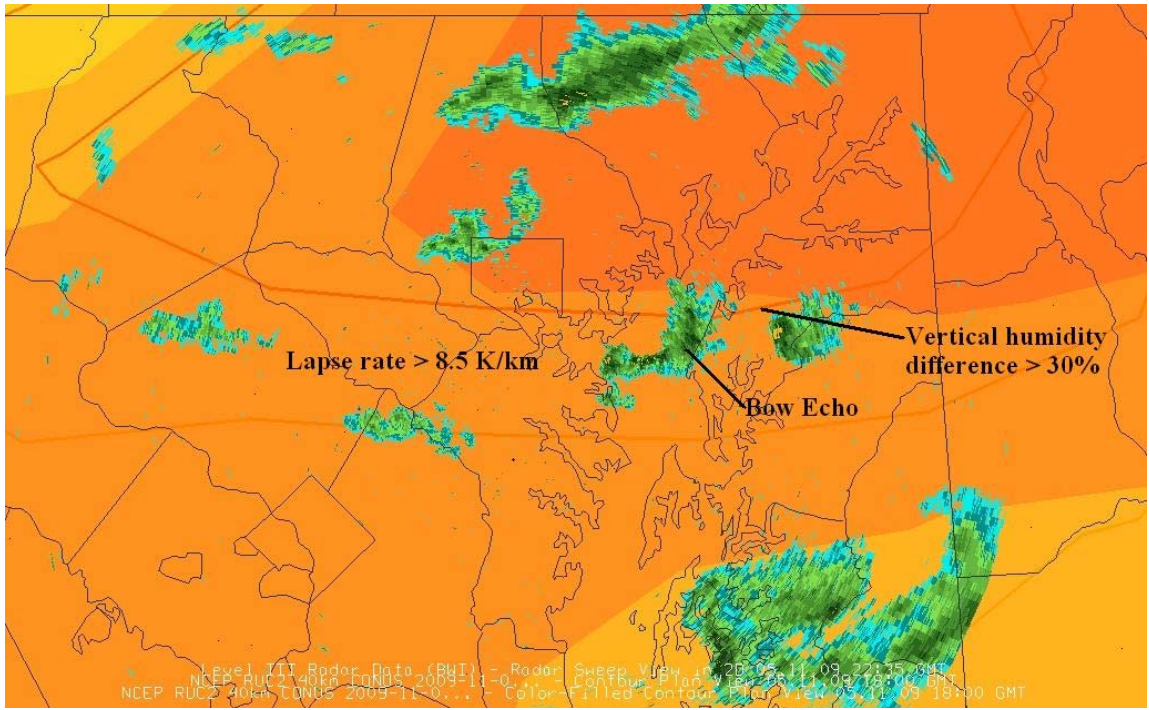

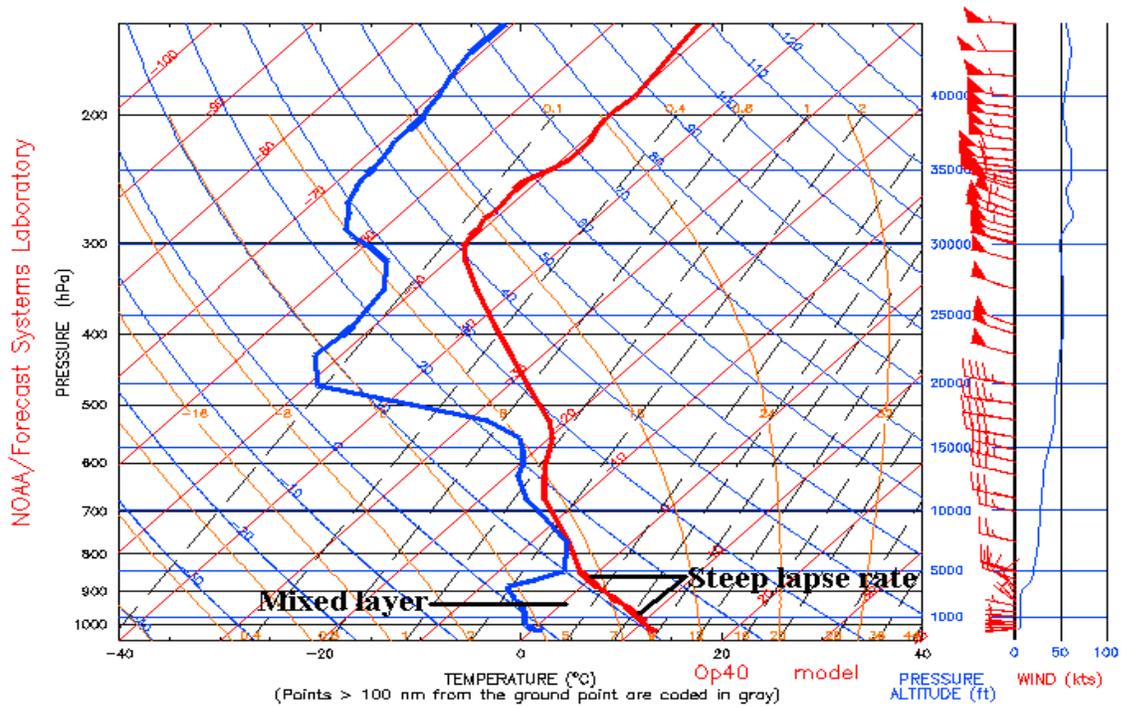

Figure 5. RUC graphical microburst product at 1800 UTC November 5, 2009 with radar reflectivity from BWI TDWR at 2235 UTC overlying the image (top) and RUC model analysis sounding over Tolchester Beach, Maryland at 1800 UTC (bottom).

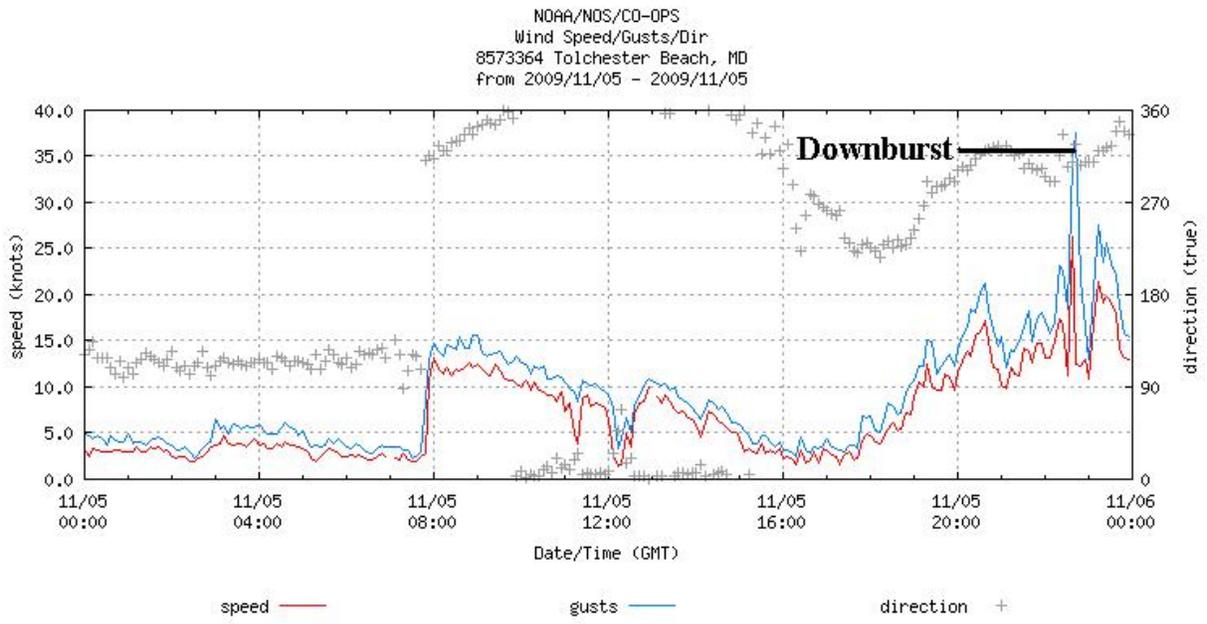

Figure 6.  Wind histogram from Tolchester Beach PORTS station.